\documentclass{article}

\usepackage{arxiv}
\newcommand{\hide}[1]{}
\usepackage[utf8]{inputenc} 
\usepackage[T1]{fontenc}    
\usepackage{hyperref} 
\usepackage{mathtools}
\usepackage{amsmath}
\usepackage{bbm}
\usepackage[export]{adjustbox}
\usepackage{multirow}
\usepackage{wrapfig}
\usepackage{graphicx}
\usepackage[export]{adjustbox}
\usepackage{xcolor}

\usepackage{subfigure}
\usepackage{paralist}
\usepackage{url}            
\usepackage{booktabs}       
\usepackage{amsfonts}       
\usepackage{nicefrac}       
\usepackage{microtype}      
\usepackage{lipsum}	
\title{Social Media Attributions in the Context of Water Crisis}

\author{
  Rupak Sarkar\\
  \texttt{rupaksarkar.cs@gmail.com} \\
   \And
Hirak Sarkar
\\
  \texttt{ hsarkar@cs.umd.edu} \\
  \And
Sayantan Mahinder \\
  \texttt{sayantan.mahinder@gmail.com} \\
  \And
Ashiqur R. KhudaBukhsh\thanks{Ashiqur R. KhudaBukhsh is the corresponding author.} \\
  \texttt{akhudabu@cs.cmu.edu} \\
}

\begin{document}
\maketitle

\begin{abstract}
Attribution of natural disasters/collective misfortune is a widely-studied political science problem. However, such studies are typically survey-centric or rely on a handful of experts to weigh in on the matter. In this paper, we explore how can we use social media data and an AI-driven approach to complement traditional surveys and automatically extract attribution factors. We focus on the most-recent Chennai water crisis which started off as a regional issue but rapidly escalated into a discussion topic with global importance following alarming water-crisis statistics. Specifically, we present a novel prediction task of \emph{attribution tie detection} which identifies the factors held responsible for the crisis (e.g., poor city planning, exploding population etc.). On a challenging data set constructed from YouTube comments (72,098 comments posted by 43,859 users on 623  relevant videos to the crisis), we present a neural classifier to extract attribution ties that achieved a reasonable performance (Accuracy: 81.34\% on attribution detection and 71.19\% on attribution resolution). 
\end{abstract}


\section{Introduction}
Water crisis is one of the pressing environmental challenges the current world is facing at present. More than a billion people do not have access to clean drinking water and nearly two million children die from water borne diseases~\cite{watkins2006human}. Groundwater, traditionally one of the main sources of water supply across the world, has taken a severe blow as one-third of the world's most extensive groundwater systems are under severe stress~\cite{richey2015quantifying}. The forecasts look even more grim; nearly two-thirds of the world population could be water stressed by 2025~\cite{seckler1999water}. While the crisis has reached an alarming level far and wide, India is listed as one of the major at-risk countries projected to have massive water scarcity in the coming decade~\cite{rost2008agricultural, postel2001dehydrating}.

In social science, apportioning attribution for a collective crisis or misfortune still remains a challenge in spite of presence of a large body of political science literature on retrospective voting(see, e.g.,~\cite{ferejohn1986incumbent, peffley1984voter}) or psychological literature on attribution (see, e.g., \cite{shaver2012attribution, schlenker1994triangle}). Moreover, in prior social science literature (see, e.g., \cite{griffin2008after, malhotra2008attributing}), attribution has primarily  relied on traditional surveys. In this paper, we explore the viability of inferring \emph{attribution ties} through large scale analysis of relevant social media discussions in the context of the Chennai water crisis. The Chennai water crisis has been a long-standing local crisis which rapidly escalated into an international talking point~\cite{Wapost2019} within the last few months revealing alarming statistics of the water crisis in India looming in near future. To the best of our knowledge, we present the first large scale social media analysis of this crisis via a substantial corpus of 72,098 YouTube comments posted by 43,859  users on 623  relevant videos. Our focus on YouTube is motivated by its global reach and popularity in the Indian subcontinent\footnote{\url{https://www.hindustantimes.com/tech/youtube-now-has-265-million-users-in-india/story-j5njXtLHZCQ0PCwb57s40O.html}}.

A thorough analysis of public opinions on a vital resource crisis such as water crisis serves multiple purposes. As discussed in~\cite{srinivasan2017prediction}, in the context of water management in developing countries, particularly in India, bridging existing communication gaps between policy-makers and the real stakeholders (people of the concerned region) can make assessing the ground realities on success/usefulness of the chosen policies (e.g., administration putting a cap on industrial water usage), quality of policy implementations (e.g., the administration passes a bill on better city planning, but no new construction is adhering to the revised guidelines), or existence of better alternatives (e.g., instead of regulating what crops can be grown, an infrastructural overhaul to promote drip irrigation may have long-term benefits) substantially easier.   

We argue that apart from complimenting traditional surveys, inferring \emph{attribution ties} from a large scale analysis of social media discussions has additional benefits. Unlike surveys, social media analyses are vastly cheaper, have faster turnaround time, can be conducted repeatedly at different spatiotemporal granularities and aggregate a larger number of opinions than traditional surveys can usually afford. For instance, the most-recent PEW survey~\cite{PewResearch} focused on India was conducted in 2018 on 2,521 users. In contrast, our data set aggregates opinions of 43,859  users.

Our Machine Learning contribution in this work is a novel task of detecting and extracting \emph{attribution ties}.
Consider the following example: \texttt{[an increased \textbf{\textcolor{blue}{population}} doesn't help but the \textbf{\textcolor{blue}{drought}} would still be there if the population hadn't increased]}. In this comment, both natural calamities and a growing population have been attributed as possible causes of the water crisis (more examples are listed in Table~\ref{tab:samples}). Our paper's main contribution is a neural classifier that automatically detects such \emph{attribution ties}. On a challenging real-world data set, our method achieved considerable accuracy in detecting presence of attribution (81.34\%) and identifying the attributing factor (71.19\%).

While aggregating opinions from social media discussions have several advantages, navigating through noisy media generated in a region of the globe with high linguistic diversity and vast number of non-native English speakers with a broad range of English proficiency poses a multitude of Natural Language Processing challenges (described in details in Section~\ref{sec:challenges}). Moreover, unlike surveys, social media comments are not always focused on the topic of discussion or a predefined goal that the survey is planned to achieve. Besides, as surveys are mostly question-answer based, there is a natural structure in the response to identify the answer from the question. Hence, both obtaining a cleaner data set focused on the topic of study and developing ML framework to extract the answers are non-trivial tasks. In addition to our ML framework to detect attribution ties, we have employed several recent advances on NLP and proposed novel data-pruning strategies to tackle these challenges.

\begin{table}[htb]
{
\small
\begin{center}
     \begin{tabular}{|p{0.26\textwidth}|p{0.60\textwidth}|}
    \hline
Attribution factor & Comment \\
\hline
Overpopulation &     \textcolor{blue}{This is a population problem}.
The earth does not have the resources to support this many people,
its reaching its carrying capacity. \\
    \hline
Climate change &   coastline cities like Mumbai and Chennai is going to sink under water after sea rise \textcolor{blue}{due to global warming} while we fight for water  \\
    \hline
Deforestation&   plant trees dumb ass
trees will hold water as well as soil
you have \textcolor{blue}{no trees} at all that is why you have not water   \\
   \hline 
Government $|$ Public water wastage &   \textcolor{blue}{not only government} but all the the civilian sorry equally responsible for or the water crisis i live in Delhi and it is not a single day when i have not encounter \textcolor{blue}{water wastage}\\
    \hline
    \end{tabular}

\end{center}
\caption{{Examples of \emph{attribution ties} in our data set. Multiple factors are separated by $|$.}}
\label{tab:samples}}
\end{table}

\section{Related Work}

Water crisis has received sustained research focus in a diverse set of fields including food policy research~\cite{hanjra2010global}, earth science~\cite{qin2007drinking}, social science~\cite{foltz2002iran}, and water research~\cite{schindler2006impending,narula2011addressing,chennai2006desalination} and encompassing a broad range of dimensions including the socio-hydrological~\cite{srinivasan2017prediction}, ethical~\cite{foltz2002iran} and cultural and  foreign policy~\cite{starr1991water} aspects of the crisis. Our work drew on these lines of research to compile a list of possible factors (listed in Table~\ref{tab:factors}) the crisis can be attributed to. However, the focus of our work is different from these lines of research as we seek to tackle the NLP challenges associated with analyzing attributions from noisy social media data. Our work shares similar motivations to a recently-reported work on the Flint water crisis~\cite{FlintBlame}. Similar to our work,~\cite{FlintBlame} constructed a substantial tweet corpus and analyzed attributions for the Flint water crisis. Our work is different from~\cite{FlintBlame} for the following two reasons. First, our data set is linguistically more challenging (challenges are outlined in details in Section~\ref{sec:challenges}) as a vast majority of the content creators are non-native speakers of English. Second, we propose an automated method to detect attributions. In contrast,~\cite{FlintBlame} formed hypotheses on the nature of the attributions and then accepted or rejected those hypotheses based on randomly sampled data labelled by annotators.  

The vast literature of political science and psychological literature do not rely on automatic methods (e.g.,  \cite{ferejohn1986incumbent, peffley1984voter,shaver2012attribution, schlenker1994triangle}). Studying attributions can be viewed as a specific case of relation extraction task~\cite{miwa-bansal-2016-end} or cause-effect analysis~\cite{hendrickx2009semeval}. In terms of motivation, our work is closely related with automatic extraction of blame ties~\cite{BlameTies2019}. Similar to~\cite{BlameTies2019}, we seek to extract causal ties between a natural resource crisis and different possible factors. However, our work differs in the following key ways. First, \cite{BlameTies2019} focused on three major US newspapers to construct their blame data set. In contrast, our focus is on social media content produced in a part of the globe where the vast majority are non-native speakers of English. Consequently, the generated texts suffer from unique challenges (described in \ref{sec:challenges}) atypical to published content in high-circulation newspapers. Second, unlike fact-based presentations in news outlets, social media discussions encompass a diverse set of expressions ranging from stating pure fact or statistics to crude disgust and subtle sarcasm. Finally, and most importantly, ~\cite{BlameTies2019} used a set of entities that are easy to detect in a sentence due to their crisp word boundaries, whereas attribution topics can be expressed in several ways. For example, both the following comments attribute \emph{deforestation} to the water crisis but have very different ways of expressing it -
\begin{compactenum}
\item \texttt{[we must protect our forests \textbf{\textcolor{blue}{plant more trees}}]}
\item \texttt{[just rewind and see how many \textbf{\textcolor{blue}{trees have vanished}} over the years to accomodate more space for buildings and malls]}
\end{compactenum}

\section{Data Set}
\subsection{YouTube Video Comments}

Using the publicly available YouTube's Search API, we queried YouTube with the following search queries: \texttt{[Chennai water crisis]}
and \texttt{[India water crisis]}. 
The choice of our first query is self-explanatory.  In order to capture the coverage of the broader focus on the country wide water crisis as the crisis unfolded and received global attention, we considered our second query. 

For each query, we constructed our video set, $\mathcal{V}$, by adding 350 recommended videos. Upon removal of duplicate videos and videos without a single comment, $\mathcal{V}$ is pruned to contain 623 unique videos. For each video in $\mathcal{V}$, we extracted posted comments using the publicly available YouTube Data API. Our overall comment data set, $\mathcal{D}_{\emph{all}}$, consists of 72,098 comments.

Since India is a country with vast linguistic diversity (22 languages recognized by the constitution among 460 languages), we observed the comments were written in a mixture of various languages either written in their native script or using Roman script.  Using a polyglot embedding based language identification technique~\cite{IndPak} that has been successfully used in another multilingual corpus generated in this part of the globe~\cite{Rohingya}, we extracted comments written in English. Our filtered set of English comments,  $\mathcal{D}$, consists of 41,791 English comments.

\subsection{Data Set Challenges}\label{sec:challenges}

While recent innovations in NLP research are increasingly capable of analyzing challenging data sets~\cite{kaufmann2010syntactic,o2010tweets, speriosu2011twitter}, much of the NLP research to date focuses on clean corpora where the content generators have sufficient proficiency in the language. In our case, the vast majority of the contributors are non-native English speakers often employing a telegraphic and colloquial style typical of social media. We outline some of these technical challenges with representative examples next.

\subsubsection{The Non-native Speaker Aspect}

\noindent \textbf{Spelling errors:} We noticed a considerable amount of phonetic spelling errors (e.g., [\texttt{check the \textbf{\textcolor{red}{expedinjar}} level in India and other countries}] originally intended to express  \textbf{\textcolor{blue}{expenditure}}).\\ 
\noindent \textbf{Code-switching:} Code-switching~\cite{thara2018code} is the phenomenon where two or more languages are mixed (e.g., \texttt{[i know bro i from maharashtra bohot buri condition hai yahapar}] translates to \emph{i know bro i from maharashtra very bad condition over here}). Our annotators observed English-Tamil, English-Hindi,  and English-Bengali code-mixed comments on our data set that made linguistic analysis more difficult.\\ 
\noindent \textbf{Out of vocabulary words:} Several comments  also used shorthand-typing dropping vowels or using non-standard abbreviations (e.g., [\texttt{\textbf{\textcolor{red}{plz}} make \textbf{\textcolor{red}{vdo}} in rainwater harvesting}]) typical to noisy social media texts, hence generating out of vocabulary words. Our data set had only 28.9\% intersection of words with GloVe~\cite{pennington-etal-2014-glove} vocabulary.\\ 
\noindent \textbf{Grammatical errors:} Since many users are non-native speakers of English, several comments suffered from grammatical disfluencies (e.g., \texttt{[this not happen everyear because of heat wave in south india this happen]}) making our analysis challenging. 

\subsubsection{Topical Focus}

Linguistic diversity and large presence of non-native speakers of English add several challenges unique to this part of the globe. However, given the geopolitical diversity of India and a multi-layered discussion on the issue with parallel threads on contemporaneous other major events, created an even bigger challenge in obtaining the subset of comments with topical focus.  


To obtain a broad overview of the topics and to demonstrate considerable presence of topics unrelated to the water crisis, we next present our topic modeling results. We considered three topic modelling algorithms: LDA~\cite{LDA} and two of it's most popular variants, MALLET\footnote{\url{http://mallet.cs.umass.edu/topics.php}} and
Biterm Topic Model (BTM)~\cite{BTM} which was used in analyzing a recent social phenomenon~\cite{demszky-etal-2019-analyzing}. As shown in Table~\ref{tab:lda} (MALLET and BTM produced qualitatively similar results), the main topics of discussion relevant to the crisis involved call to save water (topic 1), overpopulation as a major problem (topic 2), and climate change (topic 3).  
\begin{table}
  \centering
  \small
  \begin{tabular}{|c|c|c|c|c|}
    \hline
    Topic 1  & Topic 2  & Topic 3  & Topic 4 & Topic 5 \\
    (17.6\%) & (16.3\%) & (9.6\%) & (9.5\%) & (8.2\%) \\
    \hline
    water     & india    & change   & india & muslim\\
    save    & country    & climate   & pakistan & indian \\
    need    & population      & global    & river & india \\
    drink      & people & human   & china & religion \\
    river    & indian    & nature   & kashmir & hindu \\
    waste & problem  & animal & shit & like\\
    \hline
    \end{tabular}
    \vspace{0.2cm}
    \caption{{Most relevant tokens for five major topics discovered in our comments corpus $\mathcal{C}$ using~\cite{LDA}}.}
  \label{tab:lda}
\end{table}

As demonstrated in our topic modeling results, a considerable fraction of overall discussion is focused on peripheral topics unrelated to the water crisis (topic 4 and topic 5). Presence of topics surrounding India and Pakistan is not surprising since a major sociopolitical issue of India in 2019 was the Pulwama terror attack in Kashmir (14 Feb, 2019) which resulted in heightened tensions between India and Pakistan\footnote{\url{https://www.bbc.com/news/world-asia-india-47302467}} with the two countries coming almost at the brink of declaring a full-fledged war~\cite{IndPak}. We further conducted a human evaluation of randomly selected 200 comments and found that manual inspection aligns with our topic-modelling results. A small fraction of communal comments observed by our annotators align with previous finding reported in~\cite{rudra2018characterizing, rudra2016characterizing}.

\begin{figure}
\centering
\subfigure[$\mathcal{D}$]{%
\includegraphics[frame,width = 0.40 \textwidth]{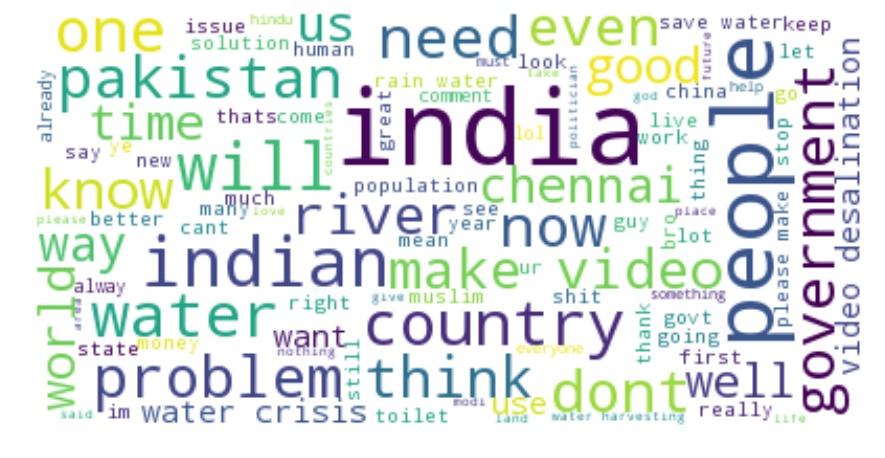}
\label{fig:local}}
\subfigure[$\mathcal{D}_{\emph{pruned}}$]{%
\includegraphics[frame,width = 0.40 \textwidth]{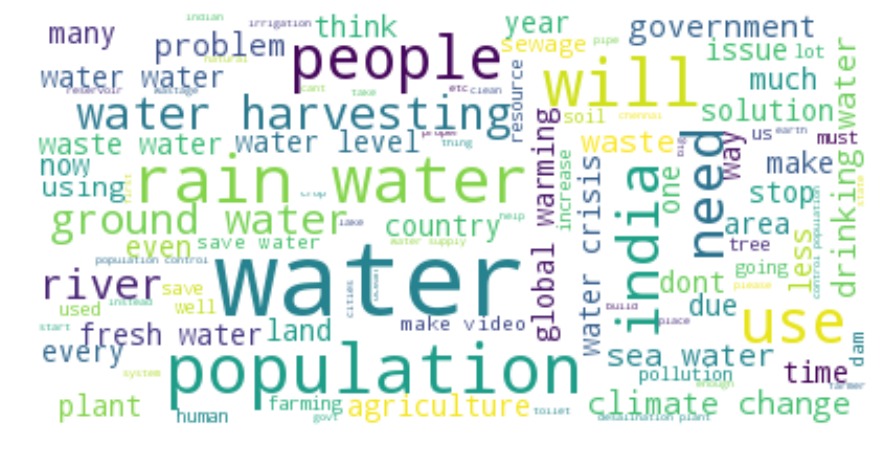}
\label{fig:all}}
\caption{\small{Word cloud visualizations of $\mathcal{D}$ and $\mathcal{D}_{\emph{pruned}}$.}}
\label{fig:pruning}
\end{figure}

\subsection{Data Pruning}
Since the peripheral discussions unrelated to the water crisis are not meaningful for our current analysis, in what follows, we describe an embedding-based pruning method to narrow down the data set with comments likelier to be relevant to the water crisis. 

\begin{table}[t]
\small
\centering
\begin{tabular}{|p{13cm}|}
\hline

topographical disadvantage, weather, climate change,  global warming, industrial development,  petroleum industry,  water intensive industries,  oil sands development
expansion of urban areas,  conversion of lands for human usage,  urban waste, corruption, mismanagement, 
contamination, industrial wastewater, industrial waste draining,  cyanobacteria,  bacteria, overpopulation,  population shift,  
excessive demand, irresponsible irrigation,  water intensive irrigation,  irrigation water demand,  irrigated agriculture,  water intensive agriculture,  inefficient irrigation,
water withdrawals, irresponsible water pumping, public water wastage,  excessive usage, indifference of policy makers, lack of funding ,  funding cuts,  lack of study,  loss of water bodies, depletion of ground water,  permanent removal from water cycle    
groundwater exploitation, strain on natural resources, deforestation, nutrient loss in soil, eutrophication, drought, flood, damming, impoundment, human activity, water intensive protein rich diet, consumption by livestock,
inefficient distribution system\\ \hline
\end{tabular}
\caption{List of factors obtained from existing water crisis literature.}
\label{tab:factors}
\end{table}

First, we ground our analysis through constructing a list of potential factors scientists typically identify as possible reasons to a water scarcity. Starting from 1990s, water crisis has received sustained attention from the water research, urban planning, political science and environmental science community. Our list (presented in Table~\ref{tab:factors}) is based on relevant literature (i) focusing on the global water crisis (ii) targeted analysis on a wide range of geographic regions, and (iii) the specific water crisis in Chennai and broadly in India~\cite{schindler2006impending,hanjra2010global, qin2007drinking, foltz2002iran,marshall2011water,rodell2009satellite,narula2011addressing,chennai2006desalination}.  

Let $\mathcal{F}$ denote the set of factors presented in Table~\ref{tab:factors}.
Let a comment $d$ be represented as a sequence of sentences $s_1,\ldots,s_n$. For each $s_i$, we compute the embedding-based cosine similarity between
$\langle s_i, f \rangle$, $f \in \mathcal{F}$ (denoted as $\emph{Cosine} (\langle s_i, f \rangle)$). We used GloVe~\cite{pennington-etal-2014-glove}, a widely-used embedding in this preprocessing step. We used the 300 dimensional GloVe model trained on 840 billion tokens of the CommonCrawl corpus, having a vocabulary size of 2.2 million. While calculating the embedding of a sentence, we removed stopwords and out of vocabulary words and computed a tf-idf weighted mean of the remaining words.
For a given comment/attribution factor pair, $\langle d, f \rangle$, the similarity score, \emph{sim}($\langle d, f \rangle$) is defined as
\emph{sim}($\langle d, f \rangle$) = max$_i$ ($\emph{Cosine} (\langle s_i, f \rangle)$))
We removed all the comments that - 
\begin{compactenum}
\item Do not fall in the top 20 percentile of the nearest comments of any attribution factor. 
\item Have a cosine similarity less than 0.7 across all factors. 
\end{compactenum}

Our pruned comment set, $\mathcal{D}_{\emph{pruned}}$, consists of 2,282 comments (9,004 sentences). A word cloud visualization (see, Figure~\ref{fig:pruning}) reveals that our pruning method lends more prominence to water specific tokens as tokens unrelated to the crisis (e.g., Pakistan) receives less prominence in the pruned data set.  We randomly sampled 1,250 comments from $\mathcal{D}_{\emph{pruned}}$ (5,004 sentences), and 1,000 comments from 
$\mathcal{D}$ (3,284 sentences) for annotation. The percentages of comments having at least one attribution from $\mathcal{D}$ and $\mathcal{D}_{\emph{pruned}}$ were 20.3\% and 88.14\% respectively. This way our embedding-based pruning method helped us collect almost four times more positive attribution data. 

Combining the samples from $\mathcal{D}$ and $\mathcal{D}_{\emph{pruned}}$, we obtained 2250 comments. This constitute the data set that we used in our prediction task. Since a comment may consist of multiple sentences with different sentence attributing to different factors, our annotators labeled at the granularity of a sentence. After annotating we found 1173 comments has at least one attribution from the set of 2,250 comments.\footnote{In order to avoid double-counting  the duplicate comments present in the sampled 1000 comments from $\mathcal{D}$, we replaced them with randomly sampled unlabeled comments to ensure the total number of labeled unique comments is 2,250.} Also at the sentence level, we obtained 2,330 positives among 8,288 total number of sentences. Since many factors listed in Table~\ref{tab:factors} are semantically close, we further define 20 broad attribution categories listed in Table~\ref{tab:categories}. 

\begin{table}[t]
{
\small
\begin{center}
     \begin{tabular}{|p{0.22\textwidth}|p{0.50\textwidth}|}
    \hline 
    Broad Category & Sub-categories \\
    \hline
Overpopulation &  overpopulation, excessive demand, population shift   \\
    \hline
Urbanization &  urbanization, expansion of urban areas, land conversion, urban waste   \\
    \hline
    Pollution &  pollution, contamination, industrial waste water, industrial draining   \\
    \hline
    Climate Change &  climate change, global warming, weather   \\
    \hline
    Agriculture &  agricultural use, water intensive irrigation, inefficient irrigation, water intensive crops   \\
    \hline
    Water Withdrawals &  water withdrawals, irresponsible water pumping   \\
    \hline
    Government Inaction &  government inaction, indifference of policy makers, lack of proper funding   \\
    \hline
    Deforestation &  deforestation, nutrient loss in soil   \\
    \hline
    Natural Calamities &  drought, flood   \\
    \hline
    Damming &  damming, impoundments  \\
    \hline
    Public Water Wastage &  public water wastage, excessive usage  \\
    \hline
    Industrial Development &  industrial development,  petroleum industry,  water intensive industries,  oil sands development
   \\
    \hline
    Corruption &  corruption, mismanagement  \\
    \hline
    Lack of infrastructure & lack of infrastructure, inefficient distribution system   \\
    \hline
    Religion &  religion, hindu caste system, islam   \\
    \hline
    Lack of Awareness & lack of awareness, lack of study   \\
    \hline
    Lack of Harvesting &  lack of rainwater harvesting, lack of water preservation  \\
    \hline
    Loss of Water Bodies & loss of water bodies, groundwater exploitation, loss of water tables   \\
    \hline
    Human activity & human activity,  water intensive protein rich diet,  consumption by livestock \\
    \hline 
    Groundwater exploitation & groundwater exploitation,  strain on natural resources    \\
    \hline
    \end{tabular}

\end{center}
\caption{20 broad categories of attribution factors.}
\label{tab:categories}}
\end{table}

\subsection{Characterizing the Annotated Data}

We had three in-house annotators proficient in Hindi, English, Bengali and Dutch. The annotation was done in two separate phases. In the first phase, the annotators only labeled if a sentence contained an attribution. The Fleiss' $\kappa$ measure of this task was high (0.87) indicating strong inter-rater agreement. For the second phase of the annotation task, our annotators had to specify the attribution factor. In this step, the annotators were allowed to discuss among themselves.  

Our annotators were instructed to label any comment with attribution and map them to the factors listed in Table~\ref{tab:factors}. While the list presented in Table~\ref{tab:factors} is comprehensive covering a broad range of countries and geographical regions, given India's multi-layered socio-political diversity, we anticipated some of the attribution factors may not be present in the compiled list. In such cases, we instructed the annotators to mark the comment as attributed and describe the category in a simple English phrase of not more than four words. For instance, religion was a category discovered by our annotators as a small fraction of comments blamed the Muslim community for overpopulation and the Hindu community for contaminating the Ganges. 

\begin{figure}[t]
\centering
\includegraphics[width=0.60\textwidth]{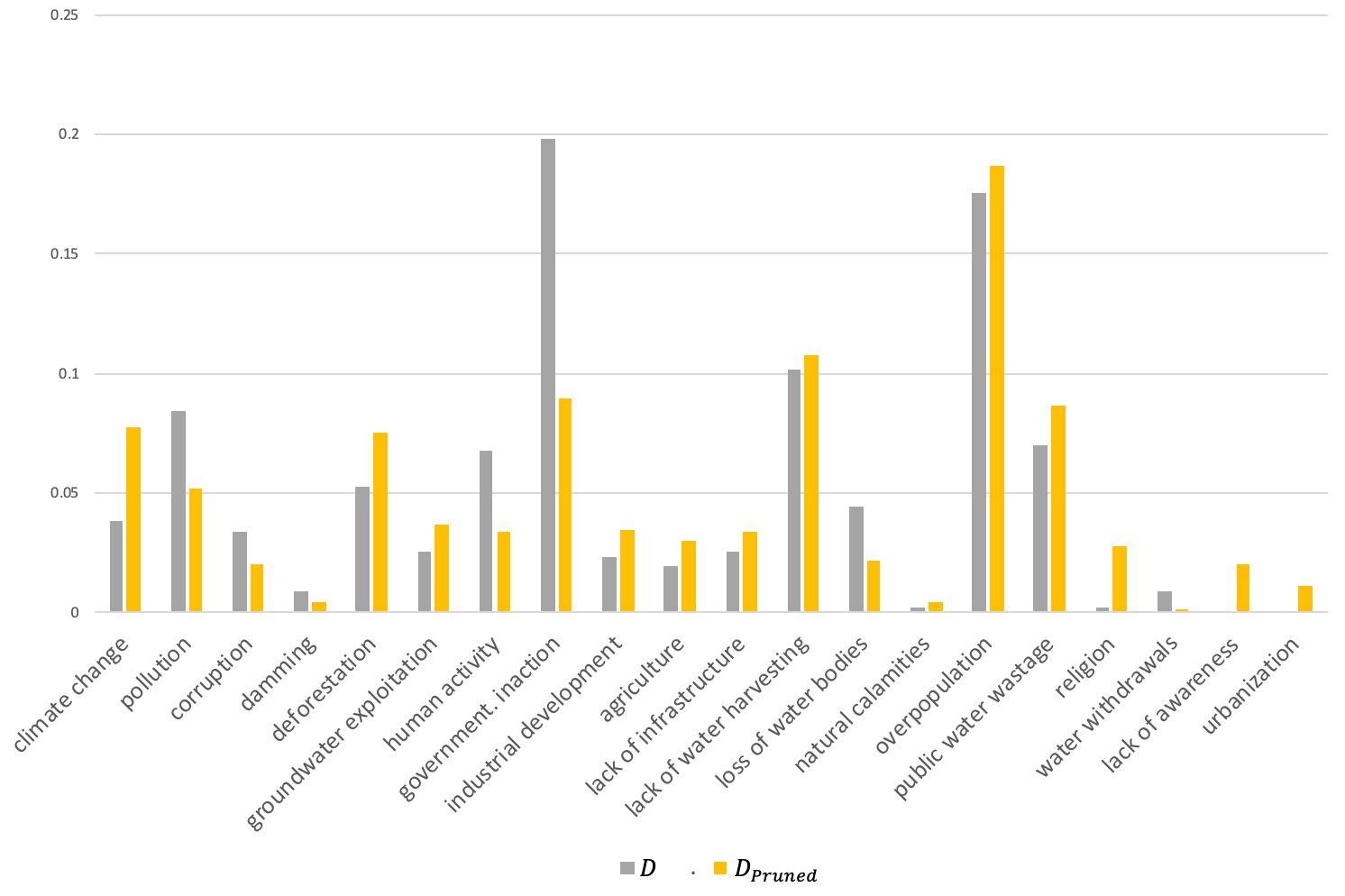}
\caption{{Breakdown of comments with attribution.}}
\label{fig:distribution}
\end{figure}

Figure~\ref{fig:distribution} presents a visualization of our comments annotated as positives. We found that overpopulation, climate change, deforestation, public water wastage, pollution and government inaction were recurrent themes in the discussion.  We now align one of the major talking points with Government policies to showcase how our analysis may provide useful data points to policy research. On a manual inspection of the pollution sub-category, we found that many comments expressed concern over open defecation and subsequent contamination of nearby water bodies. The Clean India Mission\footnote{\url{https://swachhbharatmission.gov.in/sbmcms/index.htm}} (Swachh Bharat Mission)~\cite{biswas2017swachh, bharat2016swachh} aiming to improve sanitation for millions of Indians and overall cleanliness can be viewed as a policy aligned with the aggregated public opinion in our data set.

\section{Model}

\subsection{Task}

Our prediction task is twofold: (1) attribution \emph{detection} and (2) attribution \emph{resolution}. The detection task involves predicting if a sentence contains any attribution or not. The resolution task involves correctly identifying the factor that is being attributed.

\begin{figure}
    \centering
    \includegraphics[width=0.5\textwidth]{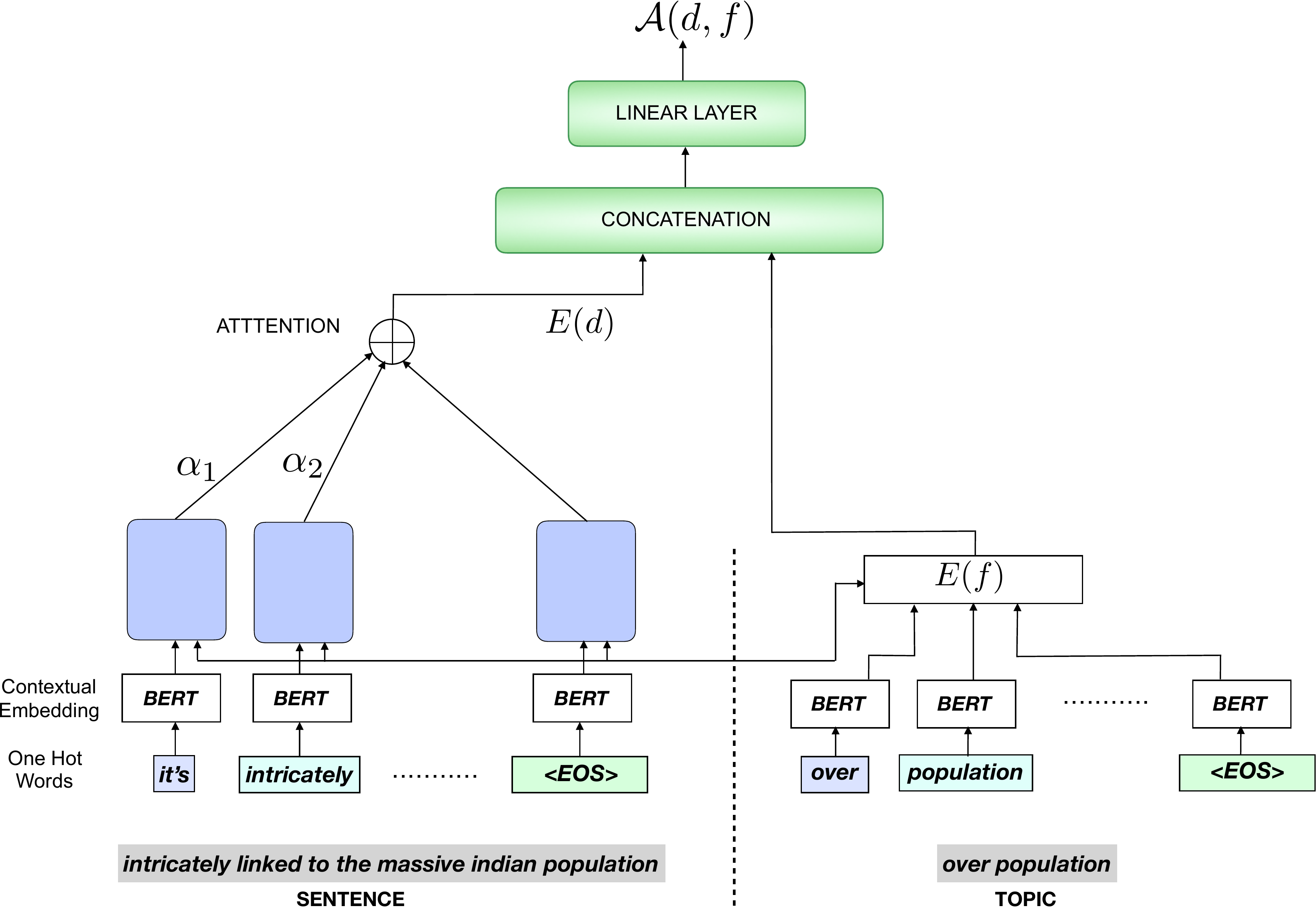}
\caption{The pipeline for training the attribution model}
\label{fig:pipeline}
\end{figure}    

\subsection{Model Architecture}

Our model architecture is illustrated in Figure~\ref{fig:pipeline}.
We aim to learn the underlying ties between a comment $d \in \mathcal{D}$ and the set of attributing factors $f \in \mathcal{F}$. However, instead of modeling the problem as a multi-class classification over $\mathcal{F}$, we modeled over a generic set of topics which is not completely known a priori. Given a set of $(d,f)$ pair where we know an attribution exists, we want to learn the different ways people comment when they attribute $f$ in a comment $d$. In order to model this generalized notion of attribution, we aim to design an attribution function $\mathcal{A} \colon \mathcal{(D, F)} \mapsto [0, 1]$ that will estimate the attribution relationship in a $(d,f)$ pair. As mentioned in previous section, we constructed our training data of $(d,f)$ pairs with of binary label $[0,1]$ to denote if an attribution $f$ is present in comment $d$. The size of our labelled data is extremely small to train a full-fledged contextual embedding model that can capture the topical relationship between $d$ and $f$. Hence, we used pre-trained contextual embeddings like ElMo~\cite{peters-etal-2018-deep} and \texttt{BERT} \cite{devlin-etal-2019-bert} to establish the relationship. Our intuition is the following: pre-trained word embeddings in a sentence capture both the semantic of the topic discussed so far as well as the context in which the topic is used. The semantic information can be used to identify the relevance of the factor $f$ in hand and the contextual information can help in identifying the use of the factor as an attribution. This intuition simplifies our learning task such that we only have to learn the factor attribution without learning the contextual representation for similarity.

The model architecture is demonstrated in Figure~\ref{fig:pipeline}. Let's assume $d = \{w_i\}$ where $w_i$ are the words in comment $d$ and similarly $f = \{w_j\}$. Using the above intuition, for a $(d,f)$ pair we first get the word embeddings $\{e(w_i)\}$ and $\{e(w_j)\}$, respectively using \texttt{BERT} or \texttt{ElMo}. As the attribution factors constitute only few words, we set the mean embedding $mean_{j \in |f|} \{e(w_j)\}$ as the representation of the factor $E(f)$. Next, using this representation $E(f)$, we determine an attribution-factor-weighted representation of comment $E(d)$ using cross-attention between the attribution factor representation $E(f)$ and the word embeddings of $e(w_i)$ from $d$. The exact formulation is as below- 
\begin{equation}
    E(f) = mean_{j \in |f|} \{e(w_j)\} \
\end{equation}
\begin{equation}
    E(d) = {\sum_{\mathclap{i \in |d|}}\alpha_{i}e(w_i)}\ where
\end{equation}
\begin{equation}
    \alpha_{i} = Cosine{(e(w_i),\ E(f))}
\end{equation}

The concatenation of the $E(d)$ and $E(f)$ is used as the final representation of the $(d,f)$ pair $E(d,f) = [E(d):E(f)]$. This is passed through a linear layer with dropouts to model the attribution function $\mathcal{A}$ and is trained with Binary Cross Entropy loss (BCELoss) using binary labels.
\begin{equation}
    A(d, f) = \sigma{(\emph{W} * E(d,f)\ + \ \emph{B})}
\end{equation}
Here $W$ and $B$ are learnable parameters and $\sigma(.)$ is the sigmoid function. 

\section{Experimental Setup}

\subsection{Model Training}
We divided our dataset into 80\% training and 20\% holdout. Further we divided the training set into 90:10 to train and model selection respectively. We used Huggingface's transformer API \cite{Wolf2019HuggingFacesTS} to build our model. We initialized our model with pre-trained weights from \texttt{BERT} (`bert-base-uncased') and trained our classifier with fine-tuning all the 12 transformer layers for two to three epochs. We used Adam optimizer \cite{journals/corr/KingmaB14} with a learning rate of 2e-5 and trained with batch size of 4. We start our model from pre-trained \texttt{ElMo} and \texttt{BERT} models and fine-tuned them along with learning the linear layer. As our training set had class imbalance, we tried to use different weights for positive samples but it didn't yield better result.  

\subsection{Performance Measures}

Since a high prediction accuracy in a task with class imbalance can be trivially achieved by predicting the majority class, we considered precision, recall and F1 score along with accuracy. For the attribution detection task, we used a threshold on the topmost attributing factor to determine if the sentence has any attribution from the provided set of factors. The threshold is tuned on the validation set mentioned in previous section. For the attribution resolution task, we used a set membership test to assign a binary outcome to a sample's prediction task. Consider $\mathcal{F}_{i}$ denotes the ground truth set of attribution factors for a sentence $s_i$, and our classifier predicts attribution factor $f_i$ for $s_i$. The binary outcome of the prediction task is  $\mathbbm{I}(f_i \in \mathcal{F}_i$) where $\mathbbm{I}$ is the indicator function. ${F}_i$ is mostly a singleton (77\% of attributed sentences have only one attribution).     

\subsection{Baseline}
We used a GloVe~\cite{pennington-etal-2014-glove} embedding based similarity measure to establish a baseline for this task. As the attribution factors can be very diverse, we chose an embedding based method. Our baseline is inspired by the observation presented in \cite{Arora2017ASB}, that weighted word embeddings produces high quality sentence representations. We used an idf (inverse document frequency) weighted sum of GloVe embedding for all the words in a sentence to get a sentence representation. Next, we used the same method for the attribution factors and computed cosine similarity to establish the relationship of the factor and the sentence.  
\section{Results}

\begin{table}
\small
\centering
\begin{tabular}{|l|l|l|l|l|}
    \hline
    & & \emph{Detection} & \emph{Resolution} & \emph{Resolution}\\
    & & & & + top 3 \\
    \hline
 \multirow{4}{*}{$\mathcal{M}$$_\texttt{BERT}$} & Precision & \textbf{64.53}  & \textbf{48.94}  & \textbf{56.10}  \\
                            & Recall 
                            & 75.72  & \textbf{39.88}  & \textbf{53.18} \\
                             & Accuracy &\textbf{81.34}  &\textbf{71.19}  &\textbf{74.96} \\
                            & F1 &\textbf{69.68}  &\textbf{43.95} &\textbf{54.60} \\
 \hline
 \multirow{4}{*}{$\mathcal{M}$$_\texttt{ElMo}$} & Precision & 54.69 & 21.09 & 31.36\\
                             & Recall 
                             &80.92 &17.92  &30.64 \\
                              & Accuracy &75.61  &57.77  &61.37 \\
                             & F1 &65.27  &19.38  &30.99 \\
 \hline

 \multirow{4}{*}{$\mathcal{M}$$_\texttt{GloVe}$} & Precision &37.60  &8.08  &13.72  \\
                             & Recall 
                             &\textbf{83.24} &12.14  &21.97 \\
                              & Accuracy &56.14  &36.01  &38.79 \\
                             & F1 &51.80  &9.70  &16.89 \\
 \hline
\end{tabular}
\caption{Performance comparison of our models and baselines. For a given task and a performance measure, the best model's performance is highlighted in bold.}
\label{tab:res}
\end{table}

\begin{table}[htb]
{
\small
\begin{center}
     \begin{tabular}{|p{0.26\textwidth}|p{0.60\textwidth}|}
    \hline
\textcolor{red}{Contamination} $|$ \textcolor{blue}{Industrial development} &     dam dont drink or consume water folks for dam whats go in must come out\\
    \hline
\textcolor{red}{Public water wastage} $|$ \textcolor{blue}{Government inaction} &   govt is not taking necessary steps to save groundwater for future\\
    \hline
\textcolor{red}{Contamination} $|$ \textcolor{blue}{Government inaction} &   the government should be ashamed for dumping sewage into the rivers   \\
   \hline 
\textcolor{red}{Public water wastage} $|$ \textcolor{blue}{Groundwater exploitation} &  we all should preserve ground water as most as we can do\\
    \hline
\textcolor{red}{Overpopulation} $|$ \textcolor{blue}{Lack of awareness} & machello very true it makes more sense to have more children children are are blessing from god there is no point talking sense to people with this mentality am so they would be very happy if all those children just dropped dead.\\
    \hline
    \end{tabular}
    
\end{center}
\caption{{Example comments that our classifier misclassified. Missclassified attribution factor is color-coded with red, ground truth is color-coded with blue.}}
\label{tab:falsenegative}}
\end{table}

\begin{table}[t]
{
\small
\begin{center}
     \begin{tabular}{|p{0.26\textwidth}|p{0.60\textwidth}|}
    \hline
Overpopulation &     stop have \textcolor{blue}{9 kids family}. \\
    \hline
Climate change &   also you should consider the extreme \textcolor{blue}{heat and lack of rain} which creates its own sanitary problems  \\
    \hline
Deforestation&   \textcolor{blue}{we cut trees} to build flat malls multi stored buildings   \\
   \hline 
Agriculture &   why is india not following the \textcolor{blue}{natural farming method} this will save water and reduce suicides\\
    \hline
Contamination &  i doubt there is one drop of \textcolor{blue}{clean water} in india with all the human and industrial waste \\
    \hline
    \end{tabular}

\end{center}
\caption{{Example comments that our classifier correctly resolved.}}
\label{tab:truepositive}}
\end{table}

Our models' performance is summarized in Table~\ref{tab:res}. $\mathcal{M}_\texttt{BERT}$, $\mathcal{M}_\texttt{ElMo}$ denote our neural classifier based on \texttt{BERT} and \texttt{ElMo} embeddings, respectively. $\mathcal{M}_\texttt{GloVe}$
 denotes our baseline. 


Since this is a task with class imbalance, precision, recall and F1 score are the more useful metrics than accuracy. We found that, on the detection task, both $\mathcal{M}_\texttt{BERT}$ and $\mathcal{M}_\texttt{ElMo}$ performed comparably. However, on the resolution task, $\mathcal{M}_\texttt{BERT}$ substantially outperformed $\mathcal{M}_\texttt{ElMo}$.

Since many of the attribution factors were semantically close (e.g., loss of water bodies, water withdrawal), we considered a more relaxed resolution criteria. In this criteria, we consider a resolution is correct if the model's top three predictions feature the ground truth. As expected, in this new metric, the performance of all algorithms improved with $\mathcal{M}_\texttt{BERT}$ still leading the pack. 
We acknowledge that our models' performance could be further improved with more sophisticated models or tuning the parameters. We are publicly releasing our data set so that the community can treat our models as baselines and to encourage further research in this domain.  

We next focus on some of the examples correctly resolved by $\mathcal{M}_\texttt{BERT}$. As shown in Table~\ref{tab:truepositive}, our $\mathcal{M}_\texttt{BERT}$ model was able to correctly identify attributions in sentences even in presence of certain degree of grammar disfluencies and an absence of the exact attribution factor per se. For instance, our model correctly resolved \texttt{[stop have \textbf{\textcolor{blue}{9 kids family}]}} to \emph{overpopulation} or \texttt{[\textbf{\textcolor{blue}{we cut trees}} to build flat malls multi stored buildings]} to \emph{deforestation} even through the specific root terms of the attribution factors were not present in the sentences. However, our model struggled when certain words are misspelled in a way that it got confused the incorrectly spelled work with another attribution factor (e.g., damn was incorrectly spelled as dam). When the attribution factors are mixed up in a confusing way (e.g., \texttt{[\textbf{\textcolor{blue}{govt is not taking}} necessary steps to save groundwater for future]}), our model failed to detect the subtleties.  We also found cases where $\mathcal{M}_\texttt{BERT}$ was able to identify the correct answers in the relaxed setting of top 3. For example,  \texttt{[we all should \textbf{\textcolor{blue}{preserve ground water}} as most as we can do]} was incorrectly attributed to \emph{Public water wastage} instead of \emph{Ground water exploitation} but the correct result was present in top 3 attributed results. We even saw in case of multiple attribution in a single sentence the model was able to identify most of the factors in top 3. Like for the sentence, \texttt{[the problem with india is their population growth corrupt government plus climate]} the model was able to detect \emph{Climate change}, \emph{Overpopulation} and \emph{Lack of awareness} as the top three attributing factors.

\section{Conclusions}

Given the preponderance of Indian population with respect to the global population (1.3 billion out of 7.7 billion), estimating awareness to climate change, growing strain on natural resources, population explosion etc. would be crucial to understand and implement better policies and any improvement in public life has the potential of positively affecting population size of global import. However, we have noticed there is a considerable gap in the NLP literature in analyzing  crises generated in the Indian sub-continent through the lens of social media, and we suspect that limitation of NLP methods to tackle the linguistic diversity and several other challenges discussed in this paper could be a possible reason for this gap. In this paper, we provide an end-to-end blue-print of analyzing social media discussions on possibly, one of the most important crises of the future: water. 

Apart from analyzing the crisis at hand, presenting promising results in a novel detection task of \emph{attribution ties}, and publicly releasing a data set in this unique domain, we consider our work fits in to the central ACL theme: \emph{Taking Stock of Where We've Been and Where We're Going} in a sense that we have employed an array of recent NLP advances~\cite{devlin-etal-2019-bert,IndPak,pennington-etal-2014-glove, peters-etal-2018-deep, BTM} to tackle a problem in a region of the globe where several events with geopolitical consequences are currently taking place~\cite{IndPak, NewYorkTimesAyodhya, WaPostKashmir}.


\bibliographystyle{unsrt}


\end{document}